\documentclass[conference]{IEEEtran}
\IEEEoverridecommandlockouts
\usepackage[left=1.5cm,right=1.5cm,top=1.5cm, bottom=1.6cm]{geometry}

\usepackage{cite, balance}
\usepackage{amsmath,amssymb,amsfonts}
\usepackage{algorithmic}
\usepackage{algorithm}
\usepackage{graphicx}
\usepackage{textcomp}
\usepackage{soul}
\usepackage{xcolor}
\usepackage{hyperref}
\usepackage[underline=true]{pgf-umlsd}
\usetikzlibrary{calc}
\usepackage{verbatim}
\usepackage{amsmath}
\usepackage{enumitem}
\usepackage{makecell}
\usepackage{mathtools}
\usepackage{amsthm}
\usepackage{tabularx}
\usepackage{pgfplots}
\pgfplotsset{compat=1.18}
\usepackage{subcaption}
\usepackage{pgfplotstable}
\usepackage{dblfloatfix}

\theoremstyle{proposition}
\newtheorem{proposition}{Proposition}
\theoremstyle{corollary}

\theoremstyle{definition}

\theoremstyle{remark}
\newtheorem{remark}{Remark}

\usepackage{caption}

\makeatletter
\newcommand{\linebreakand}{%
  \end{@IEEEauthorhalign}
  \hfill\mbox{}\par
  \mbox{}\hfill\begin{@IEEEauthorhalign}
}
\makeatother

\renewcommand{\theequation}{\arabic{equation}}

\def\BibTeX{{\rm B\kern-.05em{\sc i\kern-.025em b}\kern-.08em
    T\kern-.1667em\lower.7ex\hbox{E}\kern-.125emX}}
\begin{document}

\title{Privacy-Preserving and Simultaneous Authentication in High-Density V2X Networks

}

\author{\IEEEauthorblockN{Morteza Azmoudeh Afshar}
\IEEEauthorblockA{\textit{Informatics Institute} \\
\textit{Istanbul Technical University}\\
}
\and
\IEEEauthorblockN{Nesrine Benchoubane}
\IEEEauthorblockA{\textit{Department of Electrical Engineering} \\
\textit{Polytechnique Montr\'eal}\\
}
\and
\IEEEauthorblockN{Busra Cayoren}
\IEEEauthorblockA{\textit{Informatics Institute} \\
\textit{Istanbul Technical University}\\
}
\linebreakand
\IEEEauthorblockN{Gunes Karabulut Kurt}
\IEEEauthorblockA{\textit{Department of Electrical Engineering} \\
\textit{Polytechnique Montr\'eal}\\
}
\and
\IEEEauthorblockN{Enver Ozdemir}
\IEEEauthorblockA{\textit{Informatics Institute} \\
\textit{Istanbul Technical University}\\
}
}

\maketitle

\begin{abstract}
The rapid expansion of Vehicle-to-Everything (V2X) networks within the Internet of Vehicles (IoV) demands secure and efficient authentication to support high-speed, high-density and mobility-challenged environments. This paper presents a privacy-preserving authentication scheme that incorporates batch authentication, mutual authentication, and secure key establishment, enabling users to authenticate one another without a central authority. Our proposed scheme facilitates simultaneous multi-user authentication, significantly enhancing scalability, robustness and security in dynamic IoV networks. Results from realistic implementations show that our method achieves average authentication and verification times of 10.61 ms and 1.78 ms, respectively, for a fleet of 100 vehicles, outperforming existing methods. Scalability tests demonstrate efficient processing for larger groups of up to 500 vehicles, where average authentication times remain low, establishing our scheme as a robust solution for secure communication in IoV systems.
\end{abstract}


\section{Introduction}
Vehicular Ad-hoc Networks (VANETs) play a vital role in Vehicle-to-Everything (V2X) communication, which is a cornerstone of the Internet of Vehicles (IoV). V2X includes several communication links, such as Vehicle-to-Vehicle (V2V), Vehicle-to-Infrastructure (V2I), Vehicle-to-Pedestrian (V2P), and Vehicle-to-Network (V2N), each requiring specific access control to ensure secure and reliable communication. These links enable real-time data exchange essential for road safety, traffic management, and autonomous driving. Communication in VANETs is typically carried out through onboard units (OBUs) using WLAN-based (e.g., DSRC) or cellular-based standards (e.g., 4G, 5G, and emerging 6G) \cite{9997842, 10314795}, with external network communication managed through gateways like roadside units (RSUs) or base stations.

As VANETs evolve, the need for secure authentication becomes more pressing to prevent unauthorized access and data manipulation, which could pose significant safety risks, including cyber-attacks and traffic congestion \cite{10314795, 9279375}. Traditional authentication schemes, however, face several challenges in such dynamic environments.

\subsection{Related Work}
\label{sec:related-work}

VANET authentication has evolved to balance security, scalability, and efficiency in V2X communication, with batch and group authentication as the two main approaches.

\subsubsection{Batch authentication}
Batch authentication schemes improve efficiency by enabling simultaneous verification of multiple requests. An ID-based conditional privacy-preserving authentication (CPPA) protocol supporting batch verification without bilinear pairings is proposed in \cite{7230279}, while another approach optimizes CPPA within elliptic curve cyclic groups to reduce computational costs \cite{9372900}. A certificateless short signature-based CPPA model for efficient signature verification in V2V communications is presented in \cite{9321754}. Recent developments include a secure batch access control scheme to reduce authentication delays \cite{WANG202379} and an edge-assisted hierarchical batch authentication framework to enhance scalability \cite{10262180}. Despite these advancements, many batch authentication mechanisms still rely on centralized servers or edge nodes, creating potential single points of failure and scalability bottlenecks.

\subsubsection{Group authentication}
Group authentication schemes, such as \cite{10.1155/2021/4079092}, propose a 5G-based vehicle authentication approach using lightweight chaotic mapping for authentication and group key distribution. The scheme in \cite{9551295} enhances privacy by using self-generated pseudonyms and key pairs, eliminating the need for online group manager involvement. \cite{YA} introduces a lightweight method for generating a shared group key for machine-type communications (mMTC) without private keys, while \cite{Kubra} offers unlinkability via an anonymous public identifier but inadvertently discloses parts of the group manager’s secret and lacks support for simultaneous authentication.

Despite these advancements, several challenges remain:

\begin{enumerate}
    \item \textbf{Dependence on Group Managers:} Many schemes require the presence of a group manager during authentication, limiting autonomy and robustness;
    \item \textbf{Collusion Vulnerabilities:} Group secrets may be exposed if a sufficient number of users collaborate, enabling Sybil or collusion attacks;
    \item \textbf{Sequential Authentication:} Authentication often proceeds one-by-one, introducing delays unsuitable for dense, high-mobility V2X scenarios;
    \item \textbf{Third-Party Reliance:} Many frameworks depend on external authorities (fog nodes, edge servers) for authentication or key agreement.
\end{enumerate}

\begin{table}[htbp!]
\scriptsize
\centering
\renewcommand{\arraystretch}{1.2}
\caption{Comparison of authentication schemes in V2X networks.}
\label{tab:comparison}
\begin{tabular}{|p{0.9cm}|p{0.7cm}|p{0.7cm}|p{1.2cm}|p{0.9cm}|p{2.0cm}|}
\hline
\textbf{Scheme} & \textbf{Auth Type} & \textbf{Batch Verification} & \textbf{\makecell{Decentralized}} & \textbf{Mobility Support} & \textbf{Main Drawback} \\ \hline\hline
\cite{7230279} & Group & Yes & No & Limited & Centralized group manager needed \\ \hline
\cite{9372900} & Group & Yes & No & Moderate & Vulnerable to collusion \\ \hline
\cite{9321754} & Group & Yes & No & Limited & Heavy overhead for batch verification \\ \hline
\cite{10.1155/2021/4079092} & Group & No & No & Moderate & Requires online manager \\ \hline
\cite{9551295} & Group & No & Partial & Moderate & Scalability issues \\ \hline
\cite{YA} & Group & No & Yes & Limited & Sequential authentication \\ \hline
\cite{Kubra} & Group & No & Yes & Moderate & Risk of group secret disclosure \\ \hline
\cite{WANG202379} & Batch & Yes & No & Moderate & Third-party key dependency \\ \hline
\cite{10262180} & Batch & Yes & Partial & High & Requires edge nodes \\ \hline
\textbf{Proposed Work} & Batch + Group & Yes & Yes & High & DoS Attack \\ \hline
\end{tabular}
\end{table}

\subsection{Comparison of Existing Schemes}

Table~\ref{tab:comparison} compares representative batch and group authentication schemes based on decentralization, batch support, mobility handling, and third-party dependency. Existing methods often rely on centralized trust, lack robust mobility support, or remain vulnerable to collusion attacks. In comparison, the proposed scheme offers several key improvements:
\begin{itemize}
    \item \textbf{Decentralization:} Mutual authentication and session key establishment are achieved without the need for a centralized authority or group manager.
    \item \textbf{Collusion Resistance:} The distribution function remains hidden during key provisioning, significantly reducing the risk of collusion attacks.
    \item \textbf{Simultaneous Authentication:} Multiple participants can authenticate and establish a shared secret simultaneously, independent of group size.
    \item \textbf{Third-Party Independence:} Authentication is achieved without relying on external servers or trusted authorities, enhancing scalability and autonomy.
\end{itemize}

By addressing these fundamental limitations, the proposed scheme is better suited to the dense, high-mobility, and privacy-sensitive environments envisioned for next-generation V2X and IoV systems.

\section{Proposed Authentication Scheme}
\label{sec-method}

This section describes the proposed decentralized authentication and key establishment protocol designed for high-mobility vehicular environments.

\subsection{Setup Phase (Initial Registration)}
The initial registration phase is conducted by a group manager ($GM$), typically realized as RSU in infrastructure-supported environments. In the absence of RSU infrastructure, a trusted member device may temporarily act as $GM$. Let $GM$ be the group manager that selects a function: $$f(x),$$
where: 
\begin{itemize}
    \item The selected function is only known by the $GM$, making it the manager's secret.
    \item The function $f(x)$ must be of the form:
    \begin{equation}
        f(x) = ax + b,
    \end{equation}
    where $a, b \in \mathbb{F}_p$ are random integers,to allow for mutual authentication and $\mathbb{F}_p$ is a finite field with prime order $p$.
    \item The secret of the group manager is a polynomial $f(x)$ preferably of degree 1. The degree can be more than 1 but if the policy dictates mutual authentication in the absence of the group manager, the degree has to be 1.
    \item $GM$ defines an elliptic curve $E$ over $\mathbb{F}_p$ and chooses a generator point $P \in E(\mathbb{F}_p)$ of prime order $q$.
    \item $GM$ computes the secret group value $f(0)P$ and publishes its hash $H(f(0)P)$ using a collision-resistant hash function $H(\cdot)$ which is selected by the system administrator and known by all. 
\end{itemize}

For each member device $D_i$:
\begin{itemize}
    \item $GM$ assigns a randomly selected identifier $x_i \in \mathbb{F}_p$.
    \item $GM$ computes and securely transmits the tuple $(x_i, f(x_i)P)$ to $D_i$.
\end{itemize}

Thus, each registered device $D_i$ possesses:
\begin{itemize}
    \item A public identifier $x_i$,
    \item A private key share $f(x_i)P$,
    \item Public access to $H(f(0)P)$.
\end{itemize}

The key generation procedure is outlined in Algorithm~\ref{alg:keygen}.

\begin{algorithm}[tb]
\caption{Key Generation (Setup Phase)}
\label{alg:keygen}
\begin{algorithmic}[1]
\STATE $GM$ selects random $a, b \in \mathbb{F}_p$ and defines $f(x) = ax+b$.
\STATE $GM$ defines elliptic curve $E$ over $\mathbb{F}_p$ with generator $P$ of prime order $q$.
\STATE $GM$ computes $f(0)P$ and publicly announces $H(f(0)P)$.
\FOR{each device $D_i$ requesting registration}
    \STATE Assign random $x_i \in \mathbb{F}_p$.
    \STATE Compute $f(x_i)P$.
    \STATE Securely transmit $(x_i, f(x_i)P)$ to $D_i$.
\ENDFOR
\end{algorithmic}
\end{algorithm}

\subsection{Mutual Authentication Phase}

When members meet at the same location, they initiate decentralized group authentication without a $GM$.
\begin{itemize}
    \item Each device $D_i$ collects the public identifiers $\{x_1, x_2, \dots, x_k\}$ from each member.
    \item Each device computes its authentication contribution:
\end{itemize}

\begin{equation}
    C_i = f(x_i)P \prod_{\substack{j=1 \\ j \neq i}}^{k} \frac{-x_j}{x_i - x_j}.
    \label{eq:auth-contribution}
\end{equation}

\begin{itemize}
    \item Each device broadcasts its $C_i$ value to the others.
    \item After receiving all $C_j$ values, each device computes:
\end{itemize}

\begin{equation}
    T = \sum_{j=1}^{k} C_j.
    \label{eq:auth-aggregate}
\end{equation}

\begin{itemize}
    \item The result of this computation should be equal to $f(0)P$.
\end{itemize}

\begin{itemize}
    \item Devices verify authentication by checking:
\end{itemize}

\[
H(T) \stackrel{?}{=} H(f(0)P).
\]

\begin{itemize}
    \item If verification is successful, all devices mutually authenticate each other.
\end{itemize}

This batch authentication process is outlined in Algorithm~\ref{alg:auth}.

\begin{algorithm}[tb]
\caption{Mutual Authentication}
\label{alg:auth}
\begin{algorithmic}[1]
\FOR{each device $D_i$}
    \STATE Obtain public identifiers $\{x_1, \dots, x_k\}$.
    \STATE Compute $C_i$.
    \STATE Broadcast $C_i$ to all participants.
\ENDFOR
\FOR{each device $D_i$}
    \STATE Receive $\{C_j\}$ and compute $T = \sum_j C_j$.
    \STATE Verify $H(T) = H(f(0)P)$. 
    \IF{Verification succeeds}
        \STATE Authentication successful.
    \ELSE
        \STATE Authentication failed; abort session.
    \ENDIF
\ENDFOR
\end{algorithmic}
\end{algorithm}

\subsection{Confidential Communication Phase}

Upon successful mutual authentication, devices securely establish a common group secret $T = f(0)P$. Since $T$ is only known to authenticated members and is indistinguishable from random elliptic curve points to outsiders, it can be used for secure communications. Some bits of this value can be easily designated as the secret of the network formed by these users. Consequently, the network communication will be secure as this value is only known to the group manager and the participating members.

Potential applications include:
\begin{itemize}
    \item Deriving symmetric encryption keys from $T$,
    \item Establishing session keys by hashing parts of $T$,
    \item Authenticating subsequent group messages based on shared knowledge of $T$.
\end{itemize}

\subsection{Failure Handling}

If verification $H(T) = H(f(0)P)$ fails:
\begin{enumerate}
    
    \item Devices abort the current authentication attempt.
    \item Reorganize users into smaller groups and initiate the authentication phase for each group individually.
    \item If failures persist, return the step 2.
\end{enumerate}

\section{Security Analysis}
\label{sec:security-analysis}

The proposed authentication protocol uses Lagrange interpolation over elliptic curves, with contributions of the form:

\[
C_i = f(x_i)P \cdot \prod_{\substack{j=1 \\ j \neq i}}^{k} \frac{-x_j}{x_i - x_j}.
\]

Each participant broadcast its $C_i$, and all compute:

\[
T = \sum_{j=1}^{k} C_j.
\]

Authentication succeeds if $H(T) = H(f(0)P)$. We now evaluate the protocol’s resilience to critical attack vectors and operational failures.

\begin{proposition}
If at least one participant sends an incorrect $C_i$, the authentication fails.
\end{proposition}

\begin{proof}
Since $T = \sum C_j$ is a linear combination of all $C_i$, even a single invalid $C_j$ alters $T$, leading to $H(T) \ne H(f(0)P)$. The authentication process fails. In such a case, re-construct the group with a smaller number of participants and start the authentication process for each group. 
\end{proof}

\begin{remark}
Due to the linearity of $f(x)$, any two honest users can authenticate each other directly. This property enables smaller-sized groups to perform authentication among themselves when full group consensus fails.
\end{remark}

\begin{proposition}
A malicious node creates multiple fake identities $\{x_i'\}$ and forges corresponding $C_i$ values.
\end{proposition}

\begin{proof}
To produce valid $C_i = f(x_i')P \cdot \lambda_i$, the attacker must compute $f(x_i')P$ without knowing $f(x)$. This requires computing $f(x_i')$, which is infeasible without solving the elliptic curve discrete logarithm problem. 
\end{proof}

\begin{proposition}
Two or more legitimate users collaborate to reconstruct $f(x)$ and impersonate the group members.
\end{proposition}

\begin{proof}
Each user holds $(x_i, f(x_i)P)$, not $f(x_i)$ directly. Reconstructing $f(x)$ from elliptic curve points requires extracting $f(x_i)$, which is computationally hard due to the ECDLP. The scheme is secure against insider collusion, assuming standard cryptographic hardness.
\end{proof}

\begin{proposition}
An attacker who compromises $T = f(0)P$ at time $t$ attempts to derive $T$ values from other sessions.
\end{proposition}

\begin{proof}
If the generator point $P$ is rotated for each session, the value $f(0)P$ changes even if $f(x)$ remains the same, ensuring forward secrecy. Thus, knowledge of $T$ from past sessions does not compromise future ones.
\end{proof}

\noindent We emphasize the following remarks:

\begin{remark}
Once users employ their credentials for authentication, they should receive a new pair, as their use is safe only for a single session. Given the high mobility rate inherent to IoV entities, users are expected to be within the coverage area of distinct group managers or RSUs. 
\end{remark}

\begin{remark}
The scheme assumes all users honestly follow the protocol. If a node behaves maliciously (e.g., sending incorrect $C_i$ values), the group cannot authenticate. However, due to the ability of any two users to authenticate independently, misbehaving nodes can be excluded by subgrouping.
\end{remark}

\section{Performance Analysis}
\label{sec-simulations}

%Next, we analyze the network's performance in terms of scalability and overall network efficiency. We assess the feasibility of implementing our algorithm under real-world conditions and compare its performance against existing group authentication schemes \cite{10.1145/357980.358017, 7230279, Kubra}.  The experiments were performed on a Windows 10 machine equipped with 8GB RAM and an Intel(R) Core(TM) i7-7500U CPU @ 2.70GHz. The implementation utilized C++ with the MinGW toolchain, using g++ as the primary compiler. We employed the PARI/GP library for mathematical computations \cite{PARI2} and OpenSSL EVP with a SHA-512 equivalent digest for hashing operations \cite{Openssl}. 

% The main performance indicator used is the mutual verification time across groups of 50 to 500 vehicles. 

We evaluate the performance of the proposed scheme in terms of scalability and computational efficiency, and compare it against existing group authentication protocols~\cite{10.1145/357980.358017, 7230279, Kubra}. Experiments were conducted on a Windows~10 machine with 8GB RAM and an Intel(R) Core(TM) i7-7500U CPU @ 2.70GHz, using C++ with the MinGW toolchain and the PARI/GP library~\cite{PARI2} for elliptic curve operations. OpenSSL EVP was used for hashing with SHA-512~\cite{Openssl}.

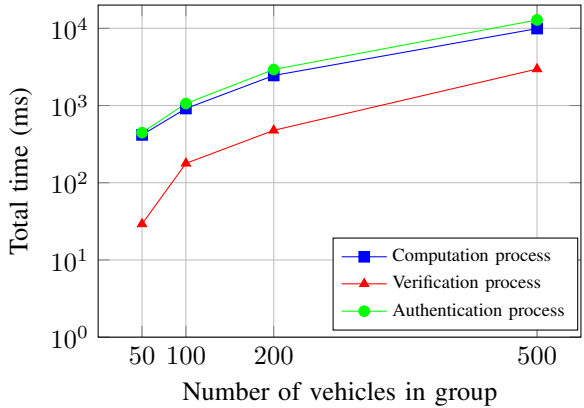
\begin{figure}[htpb!]
\centering
    \begin{tikzpicture}
    \begin{axis}[
        ylabel={Total time (ms)},
        xlabel={Number of vehicles in group},
        ylabel near ticks,
        ymode=log,
        log basis y=10,
        ymin=1, ymax=20000,
        xmin=0, xmax=550,
        xtick={50, 100, 200, 500},
        ytick={1, 10, 100, 1000, 10000},
        grid=both,
        height=6cm,
        width=8cm,
        legend style={at={(0.74, 0.3)}, anchor=north, legend cell align=left, font=\scriptsize}, % Position and left-align legend
    ]

    % Total Compute Time (Primary Y Axis)
    \addplot[color=blue, mark=square*] coordinates {
        (50, 416.247) (100, 914.277) (200, 2448.58) (500, 9917.19)
    };
    % \addlegendentry{Total computation}

    % Total Verify Time
    \addplot[color=red, mark=triangle*] coordinates {
        (50, 29.2013) (100, 178.182) (200, 476.823)  (500, 2967.72)
    };
    % \addlegendentry{Total verification}
    % Total Authentication Time (Primary Y Axis)
    \addplot[color=green, mark=*] coordinates {
        (50, 445.448) (100, 1061.01) (200, 2925.4) (500, 12841)
    };
    % \addlegendentry{Total authentication}
    \addlegendentry{Computation process}
    \addlegendentry{Verification process}
    \addlegendentry{Authentication process}
    \end{axis}
\end{tikzpicture}
\caption{Calculations of total computation, verification, and authentication times per vehicle group.}
\label{fig:multilayer-results-1}
\end{figure}

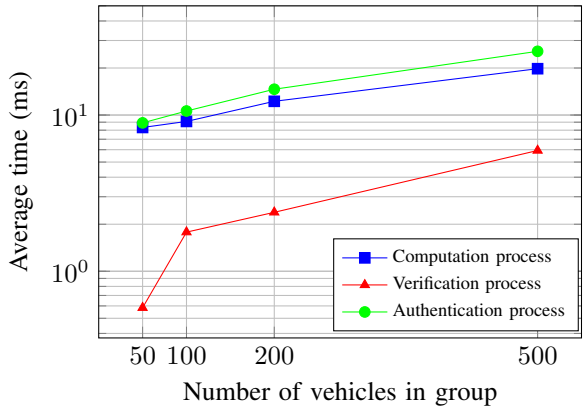
\begin{figure}[htpb!]
\centering
\begin{tikzpicture}
    \begin{axis}[
        ylabel={Average time (ms)},
        xlabel={Number of vehicles in group},
        ylabel near ticks,
        ymode=log,
        log basis y=10,
        ymin=0, ymax=50,
        xmin=0, xmax=550,
        xtick={50, 100, 200, 500},
        grid=both,
        height=6cm,
        width=8cm,
        legend style={at={(0.74, 0.3)}, anchor=north, legend cell align=left, font=\scriptsize}, % Position and left-
        % xlabel={Number of vehicles in group},
        % ylabel={Average time (ms)},
        % xmin=0, xmax=550,
        % ymin=0, ymax=30,
        % xtick={50, 100, 200, 500},
        % ytick={0, 5, 10, 15, 20, 25, 30},
        % grid=both,
        % legend style={at={(0.3, 1)}, anchor=north, legend cell align=left, font=\scriptsize}, % Position and left-align legend
        % height=6cm,
        % width=8cm,
        ]

    % Average Compute Time
    \addplot[color=blue, mark=square*] coordinates {
       (50, 8.32494) (100, 9.1) (200, 12.2429) (500, 19.8)
    };
    \addlegendentry{Computation process}
    % Average Verify Time
    \addplot[color=red, mark=triangle*] coordinates {
        (50, 0.584026) (100, 1.78) (200, 2.38412) (500, 5.93)
    };
    \addlegendentry{Verification process}
    % Average Authentication Time
    \addplot[color=green, mark=*] coordinates {
       (50, 8.90897) (100, 10.61) (200, 14.627) (500, 25.58)
    };
    \addlegendentry{Authentication process}
    \end{axis}
\end{tikzpicture}
\caption{Calculations of average computation, verification, and authentication times per vehicle group.}
\label{fig:multilayer-results-2}
\end{figure}
%We assume a one-lane road with congested traffic. In batch verification methods, we consider the average time it takes for each vehicle to perform batch verification, as this aligns with the expectation that computation will be performed on each entity. Our evaluation of computation, verification, and total authentication times demonstrates that our proposed scheme maintains efficient performance even as group sizes increase. While total computation time scales with the number of vehicles, average times remain manageable. The total computation time for a group of 50 was 416.247 milliseconds, with an average of 8.325 milliseconds. Verification took only 29.201 milliseconds, leading to a total authentication time of 445.448 milliseconds (8.909 milliseconds on average). At the maximum group size of 500, the total computation time increased to 10,067 ms (20.135 ms on average), and the verification time to 3,055 ms, yielding a total authentication time of 12,123 ms, or 26.24 ms per vehicle.

We simulate a congested one-lane road with per-vehicle computation in batch verification. As shown in Figures~\ref{fig:multilayer-results-1} and \ref{fig:multilayer-results-2}, total computation time increases linearly with group size, while per-vehicle time remains low. For 50 vehicles, total computation time is 416.2 ms, verification time is 29.2 ms, and total authentication takes 445.4 ms (8.9 ms per vehicle). At 500 vehicles, total computation rises to 10,067 ms (20.1 ms per vehicle), verification to 3,055 ms, and total authentication to 12,123 ms (26.2 ms per vehicle), demonstrating scalability.

\begin{figure}[t!]
\centering
\begin{tikzpicture}
    \begin{axis}[
        ylabel={Total mutual verification time (ms)},
        xlabel={Number of vehicles in group},
        symbolic x coords={10, 50, 100, 500},
        xtick=data,
        ymode=log, % Set y-axis to logarithmic scale
        ymin=1, % Set a minimum value to avoid log(0)
        ymax=5000000, % Adjusted max y-axis for Our method
        enlarge x limits=0.15,
        grid=both,
        height=6cm,
        width=8cm,
        legend style={at={(0.23, 1)}, anchor=north, legend cell align=left, font=\scriptsize}, % Position and left-align legend
    ]
    % Our method (line)
    \addplot[blue, thick, mark=triangle*] coordinates {(10, 2.2416) (50, 28.3043) (100, 178.182) (500, 2967.72)}; % 
    % PPA
    \addplot[brown, thick, mark=triangle*] coordinates {(10, 12.9022) (50, 463.434) (100, 1230.54) (500, 25669.7)}; % 
    % RSA
    \addplot[orange, thick, mark=triangle*] coordinates {(10, 8.348) (50, 231.935) (100,  859.688) (500, 23815)}; % 
    % ID-CCPA
    \addplot[purple, thick, mark=triangle*] coordinates {(10, 446.785) (50, 8758.67) (100, 34020.7) (500, 888359)}; %     
    \addlegendentry{Proposed Scheme}
    \addlegendentry{PPA Scheme \cite{Kubra}}
    \addlegendentry{RSA Methods \cite{10.1145/357980.358017}}
    \addlegendentry{CPPA scheme \cite{7230279}}
    \end{axis}
\end{tikzpicture}
\caption{Comparison of total mutual verification times.}
\label{fig:comparaison-1}
\end{figure}
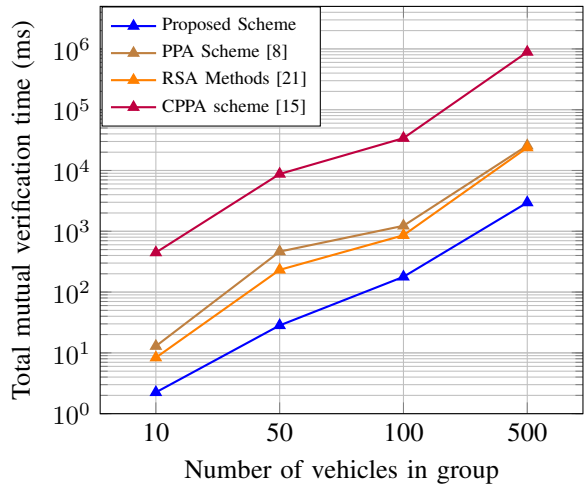

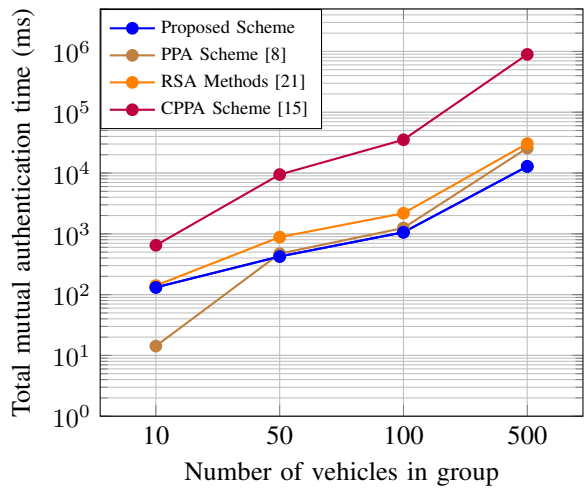
\begin{figure}[t!]
\centering
\begin{tikzpicture}
    \begin{axis}[
        ylabel={Total mutual authentication time (ms)},
        xlabel={Number of vehicles in group},
        symbolic x coords={10, 50, 100, 500},
        xtick=data,
        ymode=log, % Set y-axis to logarithmic scale
        ymin=1, % Set a minimum value to avoid log(0)
        ymax=5000000, % Adjusted max y-axis for Our method
        enlarge x limits=0.15,
        grid=both,
        height=6cm,
        width=8cm,
        legend style={at={(0.23, 1)}, anchor=north, legend cell align=left, font=\scriptsize}, % Position and left-align legend
    ]
    % Our method (line)
    \addplot[blue, thick, mark=*] coordinates {(10, 131.631) (50, 424.222) (100, 1061.01) (500, 12841)}; % 
    % PPA
    \addplot[brown, thick, mark=*] coordinates {(10, 14.2427) (50, 471.794) (100, 1248.6) (500, 25744.4)}; % 
    % RSA
    \addplot[orange, thick, mark=*] coordinates {(10, 141.047) (50, 882.475) (100, 2182.64) (500, 30519)}; % 
    % ID-CCPA
    \addplot[purple, thick, mark=*] coordinates {(10, 647.642) (50, 9442.59) (100, 35277.7) (500, 893972)}; % 
    \addplot[blue, thick, mark=*] coordinates {(10, 131.631) (50, 424.222) (100, 1061.01) (500, 12841)}; % 
    \addlegendentry{Proposed Scheme}
    \addlegendentry{PPA Scheme \cite{Kubra}}
    \addlegendentry{RSA Methods \cite{10.1145/357980.358017}}
    \addlegendentry{CPPA Scheme \cite{7230279}}
    \end{axis}
\end{tikzpicture}
\caption{Comparison of total authentication times.}
\label{fig:comparaison-2}
\end{figure}

To benchmark our scheme, we implemented the protocols in~\cite{10.1145/357980.358017, 7230279, Kubra} for a 100-vehicle cluster. As shown in Table~\ref{tab:execution-results} and Figures~\ref{fig:comparaison-1} and \ref{fig:comparaison-2}, our method achieves an average authentication time of 10.61 ms and verification time of 1.78 ms, outperforming~\cite{Kubra} by 20\% in authentication and over 7× in verification. Compared to the RSA-based scheme in~\cite{10.1145/357980.358017} (613.7 ms) and the CPPA scheme in~\cite{7230279} (352.7 ms), our protocol is up to 98\% faster.

Assuming vehicle speeds of 200~km/h, the 100--200~ms key establishment delay corresponds to a travel range of approximately 10~meters, confirming feasibility under high mobility. Our method exhibits significantly improved efficiency in both authentication and verification times. Section~\ref{sec-c-v2x} further analyzes network-level implications, including signaling overhead and latency in Cellular V2X environments using 3GPP metrics.

\begin{table}[htpb!]
    \centering
    \caption{Comparison of execution times for mutual authentication and verification for 100 vehicles.}
     \renewcommand{\arraystretch}{1.5}
    \begin{tabular}{|p{3cm}|c|c|}
        \hline
        \textbf{Methods} & \textbf{Avg. authentication} & \textbf{Avg. verification} \\ \hline\hline
        Proposed Scheme & 10.61 ms & 1.78 ms \\ \hline
        PPA Scheme \cite{Kubra} & 12.49 ms & 12.3 ms \\ \hline
        CPPA Scheme \cite{7230279} & 352.7 ms & 340 ms \\ \hline
        RSA Methods \cite{10.1145/357980.358017} & 613.7 ms & 600 ms \\ \hline
        % CPPA Scheme with Double Insurance &  &  \\ \hline
    \end{tabular}
    \label{tab:execution-results}
\end{table}

%Building on the computational efficiency demonstrated here, we extend the analysis to network-level performance in Cellular V2X environments using 3GPP metrics in Section VI.

\section{Applicability in Cellular V2X Communications} 
\label{sec-c-v2x}

Having evaluated the computational efficiency for V2V, we now extend the assessment to V2X ecosystems, examining the framework’s suitability for communication between vehicles, infrastructure, and networks using cellular technologies. Cellular V2X supports two key communication modes:

\begin{itemize}
    \item \textbf{Direct Communication (PC5 Interface)}: Allows vehicles to communicate directly with each other for safety-critical functions, such as collision avoidance and emergency alerts, without reliance on network infrastructure.
    \item \textbf{Network-Based Communication (Uu Interface)}: Connects vehicles to external networks for broader services.
\end{itemize}

The proposed scheme is compatible with both PC5 and Uu interfaces, requiring no central authority for mutual authentication or key establishment, ensuring secure communication in V2V, V2I, and V2N settings, even without a group manager.
\subsection{Analytical evaluation using 3GPP metrics}
To demonstrate our scheme’s applicability in cellular V2X, we analytically evaluate it using standard 3GPP metrics, including latency and signaling overhead. Network efficiency is assessed by profiling communication and computation overhead across the framework’s three protocol phases. Each integer (157 digits) is transmitted as 64 bytes. We report per-phase and cumulative communication overhead for $k$ vehicles.
\begin{itemize}
    \item  \textbf{Group Assignment}: $256$ bytes   (constant) $x_i, f(x_i)P, H(f(0)P)$ .
    \item \textbf{Interpolation Operation}: $(k-1) \times 64 $ bytes per vehicle.
    \item \textbf{Transmition Phase ($C_i$ values):} $(k-1) \times 128$ bytes per vehicle.
    \item \textbf{Total Overhead:} $(k-1) \times 192 + 256$ bytes per vehicle.
\end{itemize}

For 10 vehicles, the total data exchanged is 19,840 bytes, or 1,984 bytes per vehicle. Computation occurs in parallel, with per-vehicle execution time as the key metric.

% communication models for PC5 and Uu
\textbf{Communication Models:}
\begin{itemize}
    \item \textbf{PC5 (Direct Communication):} Vehicles use unicast transmissions from the GM in Phase 1 and broadcast for Phases 2 and 3.
    \item \textbf{Uu (Network-Based Communication):} Vehicles communicate through a base station, involving uplink and downlink transmissions for each phase.
\end{itemize}
% data sizes for signaling overhead

% Calculating transmission times
\textbf{Transmission Times:}
Assuming 10~Mbps for PC5 and 100~Mbps for Uu:
\begin{itemize}
    \item \textbf{For PC5}: 
    $$\frac{k*256*8}{10*(10^6)} + \frac{64*8}{10*(10^6)} + \frac{128*8}{10*(10^6)}$$
    
    \item \textbf{For Uu:} $$2*{\frac{k*256*8}{100*(10^6)}} + (2*k)*{\frac{64*8}{100*(10^6)}} + (2*k)*{\frac{128*8}{100*(10^6)}}$$
    
\end{itemize}

Computing total latency including computation times

\textbf{Latency Calculation:}
Total latency includes both transmission and average computation time from Section~\ref{sec-simulations}:
$$T_{\text{total}}= T_{\text{ph1}} +T_{\text{ph2}} +T_{\text{Ci}} + T_{\text{ph3}} $$

\begin{table}[tb!]
    \centering
    \caption{Scalability, Latency, and Signaling Overhead Results for PC5 and Uu Interfaces}
     \renewcommand{\arraystretch}{1.3}
    \begin{tabular}{|c|c|c|c|}
        \hline
        \textbf{Group Size} & \textbf{Interface} & \textbf{\thead{Total\\Latency\\(ms)}} & \textbf{\thead{Total Signaling\\Overhead\\(bytes)}} \\ \hline\hline
        10 & PC5 & 15.20 & 4,480 \\ \hline
        10 & Uu & 13.72 & 8,960 \\ \hline
        50 & PC5 & 18.46 & 22,400 \\ \hline
        50 & Uu & 11.65 & 44,800 \\ \hline
        100 & PC5 & 29.73 & 44,800 \\ \hline
        100 & Uu & 16.27 & 89,600 \\ \hline
        500 & PC5 & 122.35 & 224,000 \\ \hline
        500 & Uu & 55.64 & 448,000 \\ \hline
    \end{tabular}
    \label{tab:scalability-results}
\end{table}

As shown in Table~\ref{tab:scalability-results}, our scheme achieves latency well below the 100 ms 3GPP threshold, even for large groups. For 100 vehicles, PC5 latency is 29.73 ms and Uu latency is 16.27 ms, building on the 10.61 ms average authentication time from previous finding, confirming feasibility in high-speed environments. Our method does not affect key distribution, and computation scales linearly with vehicle count, though future work will explore how mobility factors like Doppler spread and vehicle speed impact latency and overhead.

\begin{comment}
As shown in Table~\ref{tab:scalability-results}, our scheme achieves latency well below the 100~ms 3GPP threshold, even for large group sizes. For example, at 100 vehicles, PC5 latency is 29.73~ms and Uu latency is 16.27~ms. This builds on the 10.61~ms average authentication time from Section~\ref{sec-simulations}, confirming feasibility for high-speed vehicular environments.
Our findings demonstrate that our method does not impact key distribution, and computation scales linearly with the number of vehicles. Although this analysis assumes static conditions, mobility factors such as Doppler spread and vehicle speed may affect latency and overhead, which we plan to investigate in future work.
\end{comment}

\section{Conclusion and Future Work}
\label{sec-conclusion}

\begin{comment}
This paper presents a scalable group authentication scheme that enables mutual authentication and shared key establishment without a central authority. Users securely receive keys without revealing the distribution function, allowing simultaneous authentication across different group sizes. Scalability tests show the scheme outperforms existing methods by over 20 times in larger groups. However, frequent key pair changes are required to maintain unlinkability, which introduces additional computational and communication overhead for both users and group managers.
Future research will focus on developing misbehavior-aware extensions capable of identifying inconsistent or potentially malicious actors during group authentication. Additionally, efforts will be directed toward designing dynamic rekeying protocols that intelligently adapt to evolving mobility patterns, ensuring seamless security without imposing excessive computational or network overhead on the infrastructure.
\end{comment}
This paper proposes a scalable authentication scheme that enables mutual authentication and shared key establishment without relying on a central authority. Scalability tests show over 20× improvement compared to existing methods in large groups. However, maintaining unlinkability requires frequent key pair changes, adding computational and communication overhead. Future research will focus on developing misbehavior-aware extensions capable of identifying inconsistent or potentially malicious actors during group authentication. Additionally, efforts will be directed toward designing dynamic rekeying protocols that intelligently adapt to evolving mobility patterns, ensuring seamless security without imposing excessive computational or network overhead on the infrastructure.

\bibliographystyle{IEEEtran}
\bibliography{references}

\end{document}